\documentclass[11pt]{article}

\makeatletter
\@addtoreset{equation}{section}
\def\theequation{\@arabic\c@section.\@arabic\c@equation}
\makeatother

\usepackage{amsmath}
\usepackage{amsbsy}

\makeatletter
\def\subsection{\@startsection {section}{1}{\z@}%
  {-0.5ex \@plus -1ex \@minus -.2ex}%
  {1ex \@plus.2ex}%
  {\null\normalfont\it}}
\makeatother

\newcommand{\eref}[1]{(\ref{#1})}
\newcommand{\T}[1]{\boldsymbol\Theta^{#1}}

\def\d{{\rm d}}
\def\e{{\rm e}}
\def\bi{\begin{itemize}}
\def\ei{\end{itemize}}
\def\F{{\bf F}}
\def\g{{\bf g}}
\def\u{{\bf u}}
\def\w{\wedge}

\begin{document}

\begin{center}{
{\bf On spacetimes with 3-parameter isometry group in 
string-inspired theory of gravity}\\[3mm]
P. Klep\'{a}\v{c} \\[2mm]
{\it Institute of Theoretical Physics and 
Astrophysics, Faculty of Science,\\ Masaryk 
University, Kotl\'{a}\v{r}sk\'{a} 2, 611 37  Brno, 
Czech Republic}\\[2mm]
{\small E-mail: {\tt klepac@physics.muni.cz}}\\[3mm]}\end{center}

{\small Cylindrically symmetric stationary 
spacetimes are examined in the
framework of string-inpired generalized theory of gravity. In four
dimensions this theory contains a dilatonic scalar field in addition to
gravity. A charged perfect fluid representing fermionic matter is also
considered. Explicit solution is given and a discussion of the
geometrical properties of the solutions found is carried out.\\[6mm]

\noindent Key words: cylindricall symmetry, exact solutions, 
first-order correction\\ 

\noindent MS 2000 classification: 83C22, 83C15, 83E30}

\newpage

\section{Introduction}
In this work some results on the stationary cylindrically 
symmetric spacetimes in string-inspired generalized theory 
of gravity are presented. The reason for studying this 
class of the spacetimes is twofold. First, searching for 
stringy solutions is important in itself \cite{{Barrow}, 
{Rizos}}. Second, in 
classical relativity theory the cylindrically symmetric 
spacetimes are known to violate some of the chronology 
conditions \cite{Hawking}. Therefore it is natural to 
address the question of chronology violation in the stringy 
spacetimes. Barrow and D\c{a}browski \cite{Barrow} had given 
string G\"odel-type solutions and Kanti and Vayonakis 
\cite{Kanti} found the G\"odel-type homogeneous metrics 
in the string-inspired charged gravity. 

The paper is organized as follows. In Section 
\ref{sec1} we consider an exact solution of 
a general relativistic system of a perfect 
fluid and a scalar field coupled to gravity. 
In Section \ref{sec2} the results 
of the previous Section generalized on the 
$\alpha'$-order corrections in the framework 
of the string-inspired theory \cite{{Rizos},{Kanti}}. 
Finally in Section \ref{sec3} we treat some of 
the geometrical properties of the found solution.

\section{\label{sec1} Classical solution}

Let spacetime $({\cal M},\g)$ be 
connected 4-dimensional smooth orientable manifold 
endowed with lorentzian metric $\g$.
We search for cylindrically symmetric 
stationary spacetimes. Then there exist 
local coordinate systems $(x^0,x^1,x^2,x^3)=
(t,\varphi,z,r)$ adapted to Killing fields 
$\partial_t,\ \partial_\varphi,\ \partial_z$, where 
the hypersurfaces $\varphi=0$ and $\varphi=2\pi$ are to be 
identified and $\partial_t$ is an everywhere nonvanishing 
timelike field.
 
Metric tensor field is expressed in the following way
\[\g=\eta_{\mu\nu}\T \mu\otimes\T\nu\ ,\]
where $(\eta_{\mu\nu})={\rm diag}(1,-1,-1,-1)$ is Minkowski matrix and 
$\T \mu$ are local coframe fields (greek 
indices run from 0 to 3) that form 
a (pseudo-)orthonormal basis in each cotangent space and 
are defined by 
\begin{equation}\label{basis}
\begin{array}{ll}
\T 0 = \d t+f\d\varphi\ ,  \qquad & 
\T 1 = l\d\varphi\ , \\
\T 2 = \e^\gamma\d z\ , \qquad \qquad &  
\T 3 = \e^\delta\d r\ ,
\end{array}
\end{equation}
with $f,\ l,\ \gamma,\ \delta$ being functions 
of $r$ only. Through the text the Einstein 
summation rule is used.

Our starting action is
\begin{equation}\label{clasaction}
S[\g,\phi,\mu]=\intop_{\cal M} *R-\d\phi\wedge *\d\phi
+16\pi*\mu\ ,
\end{equation}
where $R$ is the scalar curvature of the metric tensor $\bf g$, 
$\phi(r,z)$ and $\mu(r)$ are scalar fields on $\cal M$.
From the physical point of view a coupled system of 
massless scalar field $\phi(z,r)$ and perfect fluid 
is considered. 
The fluid moves in its comoving system with 
velocity vector field $\u=\T 0$ and it is 
characterised by the pressure $p(r)$ and 
the energy density $\mu(r)$. (We use the 
same symbol for vector fields and 
their naturally corresponding 1-forms.)

The equations of motion for the metric field $\g$ are the 
Einstein field equations in basis \eref{basis} written as 
\cite{Straumann}
\begin{equation}\label{Einsteinframe}
-\frac12\ {\boldsymbol\eta}_{\alpha\beta\gamma}\w
{\boldsymbol\Omega}^{\beta\gamma}=
8\pi*i_{\alpha}{\bf T}\ ,
\end{equation}
where 
\[{\boldsymbol\eta}^{\alpha\beta\gamma}=*
(\T\alpha\w\T\beta\w\T\gamma)\ ,\]
$\boldsymbol\Omega$ is curvature 2-form on $T{\cal M}$, 
\[{\boldsymbol\Omega}^\alpha_\beta=\frac 12
R^\alpha_{~\beta\gamma\delta}
\T\gamma\w\T\delta\ ,\]
and ${\bf T}$ is the stress-energy tensor
\begin{equation}\label{stress-energy}
8\pi{\bf T}=8\pi\left[(\mu+p)\u\otimes\u
-p\ \g\right]+\frac12\left[
\d \phi\otimes\d\phi-\frac12\ \g(\d\phi,\d\phi)\ \g\right]\ .
\end{equation}
The appropriate explicit form of the Einstein 
equations for the basis \eref{basis} is
similar to the one in \cite{klep} so we 
simply refer the reader to the paper.

Bianchi identity ${\rm D}*i_\alpha{\bf T}=0$ in our case 
implies $p={\rm const}$ because of the fluid 
particles geodesic motion (see below).

Altogether we obtain five independent equations for six 
unknown functions: $f,\ l,\ \gamma ,\ \delta ,\ \mu$ 
and $\phi$. Therefore the mathematical problem admits 
one degree of freedom corresponding to the $r$-coordinate 
rescaling possibility as can be seen directly from 
\eref{basis}.

The scalar field equation of motion is the 
massless Klein-Gordon equation
\begin{equation}\label{Klein-Gordon}
*\ \d*\d\ \phi=0\ .
\end{equation}
Because of the independence of $r$ and $z$ coordinates 
one can carry out the separation of variables in 
\eref{Klein-Gordon} to obtain
\begin{equation}\label{dilaton}
\phi = \phi_0+\phi_1z+\phi_2\int l\ \e^{\delta-\gamma}\d r\,,
\end{equation}
where $\phi_0,\ \phi_1$ and $\phi_2$ are constants 
such that $\phi_1\cdot\phi_2=0$ .

For the function $l^2$ one gets a second-order non-linear 
equation that can be transformed into the following form
\begin{equation}\label{non-lin}
\frac\d{\d x}\left[\frac{y'+x}y\right]=
-\frac {k\phi_2^{~2}}{y^2}\ ,\qquad k= {\rm const}\ .
\end{equation}
\def\o{\otimes}

We proceed further by dividing solutions 
of \eref{non-lin} into two groups according as 
$\phi_2$ is zero ({\it Case (a)}) or is not 
({\it Case (b)}).

\subsection*{Case (a): $\phi_2=0$}

From \eref{dilaton} we can see that $\phi$ depends linearly 
on $z$ alone. The solution of the Einstein equations can be 
settled in the form 
\begin{eqnarray}
 {\bf g}=\d t\ \o\ \d t+\frac\Omega C\ \gamma\ (\d t\ \o\ \d \varphi+
\d \varphi\ \o\ \d t)-l^2\d\varphi\ \o\ \d\varphi\nonumber\\
-\ \e^{2\gamma}
\d z\ \o\ \d z-C^{-2}l^{-2}\d\e^\gamma\ \otimes\ \d\e^\gamma\ . 
\label{classolution}\end{eqnarray}
The metric function $l^2$ is given by  
\begin{equation}\label{l^2}
l^2=\frac{8\pi p}{C^2}\ \e^{2\gamma}-
\frac{4\Omega^2+\phi_1^2}{2C^2}\ \gamma+\nu\ ,
\end{equation}
with $\Omega,\ \nu,\ C$ as integration constants and
$\gamma$ an arbitrary non-constant $C^2$ function. 

\subsection*{Case (b): $\phi_2\neq0$}

Now we have dust distribution ($p=0$) that generalizes the 
van Stockum solution and the explicit form of the 
metric is omitted here for the reason given below. 
\newcommand{\La}[1]{\left(a{\tilde \gamma}+b\right)^{#1}}
The scalar field is integrated to
\[\e^{\phi}=\La{\frac{\phi_2}{a}} ,\]
with ${\tilde\gamma}$ being an arbitrary function and 
$a$ and $b$ constants.\\

In this work we focus on the spacetimes with 
3-parameter isometry group, especially on the 
stationary cylindrically symmetric ones. 
If one requires the found solutions to be cylindrically symmetric 
regular at the origin the axial symmetry condition
% \begin{equation}
%X\equiv {\bf g}(\partial_\varphi,\partial_\varphi)
%\propto {\cal O} (r^2)\quad {\rm as}\ r\rightarrow +0\label{axial}
%\end{equation}
and the elementary flatness condition 
%\begin{equation}
% \displaystyle\lim_{r\rightarrow +0}\frac{{\bf g}
%(\nabla X,\nabla X)}{4X}=1\label{flatness}
%\end{equation}
have to be imposed \cite{Kramer}. 
For {\it  Case (a)} these conditions yield 
\[8\pi p-\Omega^2-\frac14 \phi^2_1=C, \quad 
8\pi p+C^2\nu=0\ .\]

\noindent On the hand it turns out that {\it Case (b)} 
cannot be cylindrically symmetric.\\

\noindent For further investigation of the charged scalar
field and allowed frequencies in G\"odel-type
background we refer the reader to \cite{Radu},
where is also considered the problem of field
quatization using the Euclean
approach to quantum field theory.

\section{\label{sec2} String-inspired theory}

\newcommand{\Om}{{\boldsymbol\Omega}}
\def\A{{\bf A}}

We consider a generalized theory of gravity which describes 
the coupling of a dilatonic scalar field to an electromagnetic 
field and gravity with the following $\alpha'$-order corrected 
effective action in the Einstein frame 
\cite{Kanti} 
\begin{eqnarray}
  S_{\rm eff}[{\bf g},\A,\phi,\mu]=\intop_{\cal M}\left[*R
+16\pi*\mu-\d\phi\wedge *\d\phi+2\lambda\ \e^{\phi}\ \F\wedge*\F
\right]\ . \label{stringaction} 
\end{eqnarray}
Here $\F$ is the 
electromagnetic field 2-form. Coupling 
constant $\lambda$ is expressed like 
$\lambda=\frac{\alpha'}{4g^2}$, where 
$\alpha'$ is the inverse string tension 
(Regge slope) and $g$ is essentially the 
string coupling constant. Physically, if 
the fluid particles are charge carriers, 
the charged perfect fluid is rough 
approximation of a fermionic matter 
in the theory.

We are interested in $\alpha'$-order 
corrections to the classical solution 
\eref{classolution}. It means that in 
\eref{stringaction} we keep only zeroth 
and first order terms in $\alpha'$. 

The scalar field $\phi$ is written like $\phi=\phi^{(0)}+
\alpha'\phi^{(1)}$, where $\phi^{(0)}$ is a classical 
zero-order solution. As {\it Case (b)} of Section 
\ref{sec1} cannot 
represent cylindrically symmetric spacetime, we take 
\[\phi^{(0)}=\phi_0+\phi_1z\ .\]

\noindent It should be mentioned that 
in some works \cite{Rizos} the additional 
term 
\[S_{\rm GB}=8\pi^2\lambda\intop_{\cal M}\e^\phi e({\cal M})\ ,\]
enters action \eref{stringaction}, 
where $e({\cal M})$ is the Euler class 
of the manifold $\cal M$, in four-dimensions 
equal to  
\[e({\cal M})=\frac 1{8\pi^2}
\left(R_{\alpha\beta\gamma\delta}R^{\alpha\beta\gamma\delta}-
4R_{\alpha\beta}R^{\alpha\beta}+R^2\right){\boldsymbol\eta}\ ,\]
$\boldsymbol\eta$ is the volume element. 

Next analysis will be simplified if one 
adopts the following reasonable conditions 
on the electromagnetic field behaviour.
First, the Lorentz force $\cal F$ acting 
on the fluid particles, 
${\cal F}\propto*\left(\u\wedge *\F\right)$, 
vanishes (for the case with the non-vanishing 
Lorentz force but without the scalar field 
see \cite{klep}). Second, only longitudinal 
magnetic field survives. In comoving system the 
magnetic field 1-form is given by ${\bf B}=
*(\u\w\F)$, which together with the previous 
point gives 
\[*(\u\w \F)\w \T 2=0\ .\]
 
Before treating the equations of motion for 
the electromagnetic and gravitational fields 
we impose another supposition concerning the 
perfect fluid. 

From the physical considerations it 
follows, because the fluid represents the fermionic 
matter, that it is more physically favourable if one 
assumes the presence of a continuous electric 
charge distribution throughout the spacetime 
with a charge current density 1-form $\bf j$. 
Nevertheless, to maintain the action 
\eref{stringaction} unaffected, the current 
density incorporation is permissible only in 
the case where the corresponding source term 
${\bf  j}\w*\A$ does not act as an extra 
gravitational field source. 

Furthermore it is natural to postulate that 
the fluid particles are charge carriers. Thus 
the current density $\bf j$ is purely convectional, 
${\bf j}=\rho\u$, with $ \rho(r)$ being an 
invariant charge density.  

With the above in mind the generalized 
Maxwell equations take the form
\begin{equation}\label{Maxwell}
-*\d* (\e^{\phi^{(0)}}\ {\bf F})=\frac{4\pi}\lambda\rho\ \T 0\ .
\end{equation}

By virtue of the Einstein-Maxwell equations 
resulting from \eref{stringaction} (see below), 
the conditions above and \eref{Maxwell} will 
be satisfied if 
\begin{equation}\label{elmag}
{\bf F}=B\exp(-\frac12\phi_0-\frac12\phi_1z
-\gamma)\ \T 3\w\T 1\ , \qquad B={\rm const}\ .
\end{equation}
Note that 
\begin{equation}\label{monopol}
*\d \F=\frac12B\phi_1\exp(-\phi^{(0)}/2-2\gamma)\ \T 0\ ,
\end{equation}
which signals that we deal with a magnetic 
monopoles current. As a matter of fact it 
is neccessary to introduce a magnetic charges 
current density 1-form ${\bf j}_m$. In 
principle there are possibilities to retain 
\eref{monopol} physically admissible. Either 
we can expect that going to a non-abelian gauge 
fields will smooth out this solution, or, in 
the case of abelian gauge fields, it is possible 
to introduce ${\bf j}_m$ explicitly in the action 
\eref{stringaction}, but one has to break the general 
covariance to do this \cite{Schwarz}.

The metric field equations of motion 
following from the superstring effective 
action \eref{stringaction} are the Einstein 
equations \eref{Einsteinframe} enriched by the 
electromagnetic field contribution ~\cite{Straumann} 
\[8\pi{\bf T}_{\rm elmag}=\lambda\frac{\e^{\phi^{(0)}}}{8\pi}
\ \T \alpha\otimes*
(\F\w i_\alpha*\F-i_\alpha\ \F\w*\F)\] 
to the stress-energy tensor field \eref{stress-energy}. 
Again we refer the reader to \cite{klep} for 
details.

One finds that one has six independent equations 
for totality of seven unknowns: 
four metric functions $f,\ l,\ \gamma,\ \delta$ 
and three physical quantities $ \phi,\ \mu,\ \rho$. 
Again, we have one degree of freedom due to 
the $r$-coordinate rescaling possibility.

Bianchi identity in our case, provided 
that the scalar field equation of 
motion is fulfilled, is
\begin{equation}\label{Bianchi}
\u\w*(\mu\ \d {\bf u}+\d(p\ \u)+\lambda\ \rho\ {\e^{\phi^{(0)}}}\ \F)=0\ .
\end{equation}
Inserting \eref{basis} and \eref{elmag} 
into \eref{Bianchi} one finds that the 
pressure $p$ is constant. 

The mathematical structure of the 
Einstein-Maxwell equations is much the 
same as classical theory, Section \ref{sec1} 
(see also \cite{klep}). As a consequence, 
if one requires solution to be cylindrically 
symmetric, the scalar field becomes equal to
\[\phi=\phi_0+{\tilde\phi}_1z\ ,\]
where ${\tilde\phi}_1$ is constant.

Resulting metric is given by
\begin{eqnarray} 
  \g  =\d t\otimes\d t+\frac {\Omega}{C} \gamma\ 
(\d t\ \otimes\ \d \varphi+\d \varphi\ \otimes\ \d t)
-l^2\d\varphi\ \otimes\ \d\varphi\nonumber\\
-\ \e^{2\gamma}\ 
\d z\ \otimes\ \d z-C^{-2}l^{-2}\d\e^\gamma\ \otimes\ \d\e^\gamma\ ,
\label{stringsolution}\\[3mm]
l^2=\frac{8\pi p}{C^2}\ \e^{2\gamma}
-\frac{4\Omega^2
+{\tilde\phi}_1^{~2}-4\lambda B^2}{2C^2}\, \gamma
+\nu\ .\nonumber
\end{eqnarray}
Found solution \eref{stringsolution} is very 
similar to the zero-order solution \eref{classolution}. 
The only difference appears in $l^2$ function 
\eref{l^2}, namely the coefficient of the 
linear term in $\gamma$ gets shifted due to 
the magnetic field presence. 

The formula for the energy density 
\[\mu=\frac1{4\pi}\left(2\Omega^2-\lambda B^2\right)\e^{-2\gamma}-3p\]
has very transparent physical interpretation. 
A ``specific" mass density $\mu+3p$ must be 
added to the magnetic energy density to 
balance the rotation.

As an important example of \eref{stringsolution} 
we choose the integration constants like 
\begin{equation}
 16\pi p=-C^2\nu =m^2,\ \gamma=\frac{2C}{m^2}\ {\rm sh}
\left(\frac {mr}2\right),\ 
4\Omega ^2=4 \lambda B^2-{\tilde\phi} _1^2+2m^2(1-C)\ ,\label{constants}
\end{equation}
where $m$ is constant. With this choice the 
energy density becomes equal to
 \begin{equation}
 8\pi\mu=\left(2 \lambda B^2- {\tilde\phi}_1^2+
2m^2[1-C]\right)\e^{-2 \gamma}-
\frac{3m^2}2\ .\label{genkantidens}
\end{equation}
In \eref{constants} and \eref{genkantidens} the 
substitution $Cm^{-2}\rightarrow C$ is assumed. 
Since the pressure is constant it follows 
that our spacetime can be reinterpreted as 
the charged dust solution (with the vanishing 
pressure) and the non-zero cosmological constant 
\cite{klep}. 

Note that if one wants at this stage 
to eliminate the perfect fluid contribution 
to the action \eref{stringaction} and consider 
only coupling of the electromagnetic and 
the dilatonic fields to gravity in a spacetime 
with the non-zero cosmological constant 
the following equality must hold
\begin{equation}
\mu+p=0\ .\label{nofluid}
\end{equation}
In this case the equations \eref{genkantidens} 
and \eref{nofluid} imply $C=0$ along with 
${\tilde\phi}^2_1=2 \lambda B^2+m^2$. The metric 
tensor is found to be
\begin{eqnarray}
\g =\left(\d t+\frac{4 \Omega}{m^2}\ {\rm sh}
\left(\frac{mr}2\right)
\d\varphi\right)\otimes 
\left(\d t+\frac{4 \Omega}{m^2}\ {\rm sh^2}
\left(\frac{mr}2\right) \d\varphi\right)\nonumber
\\-\frac1{m^2}\ {\rm sh}^2(mr)\ \d\varphi
\otimes\d\varphi-\d z\otimes\d z-\d r\otimes\d r\ ,
\label{kanti}
\end{eqnarray}
with $ \Omega $ subject to $4 \Omega^2=2 \lambda B^2
+m^2$. The dilaton is approximately given by
\begin{equation}\label{kantidilaton}
 \phi = \phi _0+\left(\phi _1+\frac{ \lambda B^2}m\right)z\ .
\end{equation}

The solution \eref{kanti} and \eref{kantidilaton} 
manifestly describes the G\"odel-type spacetime 
\cite{Reboucas} and was found in \cite{Kanti} 
by another way when studying a homogeneous 
G\"odel-type solutions. Note that in zero-order regime 
$ \lambda\rightarrow 0$ one has $ \phi^2_1=m^2$ and 
$4 \Omega^2=m^2$, the latter equality immediately implying 
$g_{\varphi\varphi}=\partial_\varphi\cdot \partial_\varphi\leq0$.  
Thus there are no closed timelike curves 
in the spacetime \cite{Hawking}. On the other hand in 
$ \alpha'$-order framework $g_{\varphi\varphi}$ 
becomes positive for sufficiently large $r$ 
(because $ \lambda$ is positive). In this way 
the first-order correction causes the chronology 
violation. 
 
\section{\label{sec3} Geometrical properties}

In each tangent space $T_p\cal M$ the projection 
tensor onto 3-dimensional subspaces $W_p$, 
orthogonal to $\u$, is given by
\[{\bf h}={\bf g}-\u\otimes \u\ .\]
The tensor field $-\bf h$ serves a positive 
definite metric on $W_p$. 
 
For a given spacetime to be static there 
must exist hypersurface-orthogonal timelike 
Killing vector field \cite{Kramer}. In our 
case the vorticity 1-form $\boldsymbol\omega$ 
equals
\[{\boldsymbol\omega}=\frac12*(\u\w\d \u)=\Omega\d z\ .\] 
In this way the metric \eref{stringsolution} 
is static if and only if $\Omega=0$. Geometrically, 
a collection $W=\displaystyle\cup_pW_p$ is not involutive.

The fluid particles acceleration $\dot\u$ is given by 
\[\dot \u=-*(\u\w*\d \u)\ ,\]
and vanishes for \eref{basis} showing that $\u$ parallelly 
propagates along itself and fluid particles move geodesically. 
Namely for this reason, and because we have restricted our 
attention to the Lorentz force-free case, the pressure has 
to be constant. 

The rate-of-strain tensor field is actually the extrinsic 
curvature $\bf K$, and is defined by
\def\Lu{\mathop\pounds_\u}
\begin{displaymath}
{\bf K}=\frac12\Lu {\bf h}\ .
\end{displaymath}
Direct computation shows that $\bf K$ vanishes identically 
for \eref{stringsolution} which from the physical viewpoint 
means that the fluid rotates as a rigid body.

The last remark concernes the algebraic 
classification of the Weyl tensor field 
$\bf C$. It turns out that for \eref{stringsolution} 
there generaly exist just four distinct 
null vectors $\bf k$ (modulo multiplying 
${\bf k}\rightarrow a{\bf k}$, $a$ is constant) 
satisfying 
\[k^{\beta} k^{\gamma} k_{{\lambda}}
C_{\alpha\beta\gamma\delta}
k_{\sigma}(\T\lambda\w\T\alpha)
\otimes(\T\delta\w\T \sigma)=0\ .\]
Thus our spacetime \eref{stringsolution} belongs to 
the type $I$ according to the Petrov classification 
\cite{Kramer}. 

\section*{Acknowlegments}
The author is obligated to Dr Rikard von Unge for 
helpful disscussions.

\end{document}